\documentclass[11pt]{report}

\usepackage[dvipdfm]{graphicx}
\usepackage{amsmath}
\usepackage{bm}
\usepackage{fancyhdr}
\usepackage{makeidx}

\newcommand\bk{{\bf k}}

\newcommand\bp{{\bf p}}
\newcommand\br{{\bf r}}

\newcommand\nn{\nonumber \\}
\newcommand\beq{\begin{equation}}
\newcommand\eeq{\end{equation}}
\newcommand\beqa{\begin{eqnarray}}
\newcommand\eeqa{\end{eqnarray}}

\newcommand{\ds}[1]{#1 \hspace{-0.5em}/}  
\newcommand\bzeta{\mbox{\boldmath$\zeta$}}
\newcommand\bB{\mbox{\boldmath$B$}}
\newcommand\bgamma{\mbox{\boldmath$\gamma$}}
\newcommand\btau{\mbox{\boldmath$\tau$}}

\newcommand\bq{{\bf q}}

\newcommand{\lk}{\left(}
\newcommand{\rk}{\right)}

\newcommand{\ldk}{\left[}

\newcommand{\rdk}{\right]}

\newcommand{\va}{\mbox{$\bm{a}$}}
\newcommand{\vp}{\mbox{$\bm{p}$}}
\newcommand{\vq}{\mbox{$\bm{q}$}}
\newcommand{\vr}{\mbox{$\bm{r}$}}
\newcommand{\vk}{\mbox{$\bm{k}$}}

\makeatletter
\def\ps@fancy{%
\def\chaptermark##1{\markboth{\ifnum \c@secnumdepth>\z@
Chapter \thechapter\hskip 1em\relax \fi ##1}{}}%
\def\sectionmark##1{\markright{\ifnum \c@secnumdepth>\z@
\thesection\hskip 1em\relax \fi ##1}{}}%
\def\subsectionmark##1{\markright {\ifnum \c@secnumdepth >\@ne
\thesubsection\hskip 1em\relax \fi ##1}}%
\ps@@fancy
\gdef\ps@fancy{\@fancyplainfalse\ps@@fancy}%
\ifdim\headwidth<0sp
\global\advance\headwidth123456789sp\global\advance\headwidth\textwidth
\fi
}
\makeatother

\makeindex

\begin{document}
\baselineskip=16.0pt
\newcounter{Mycounter}\setcounter{Mycounter}{\value{chapter}}
\addtocounter{chapter}{9}

\chapter{
Magnetic field and quark matter in the core
}

\vspace*{-0.5cm}
\centerline{
Toshitaka Tatsumi{\footnote{E-mail: tatsumi@ruby.scphys.kyoto-u.ac.jp}}
}
\centerline{
Department of Physics, Kyoto University, Kyoto 606-8502
}

\

Magnetic properties of quark matter are discussed in the light of the observation of pulsars. 
Our works about spontaneous spin polarization and spin density wave are reviewed and 
their implications on compact-star phenomena are discussed. 
In particular, the former subject may be directly related to the origin of strong magnetic fields. 
An inhomogeneous state emerges following the chiral transition, 
where a kind of spin density wave develops.

\section{Introduction}

Pulsars are rotating neutron stars and have a strong magnetic field, by which they emit optical, radio , X-ray and $\gamma$-ray pulses.

\begin{figure}[h]
\begin{minipage}{0.4\textwidth}
\begin{center}
\includegraphics[width=4.8cm]{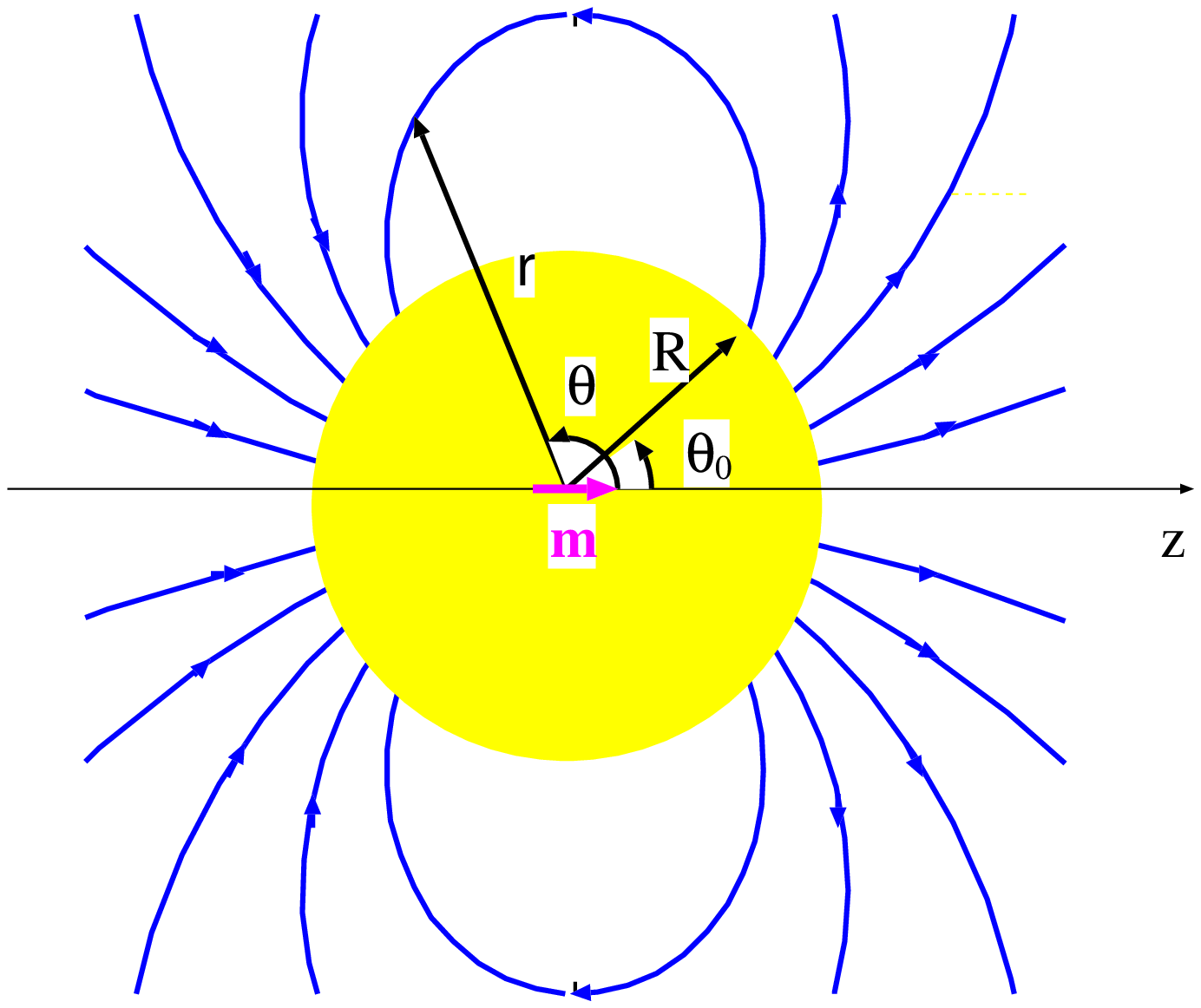}
\end{center}
\caption{Dipole magnetic field-lines outside the star.}
\label{dipole}
\end{minipage}
\hspace{\fill}
\begin{minipage}{0.56\textwidth}
\begin{center}
\includegraphics[width=4.5cm]{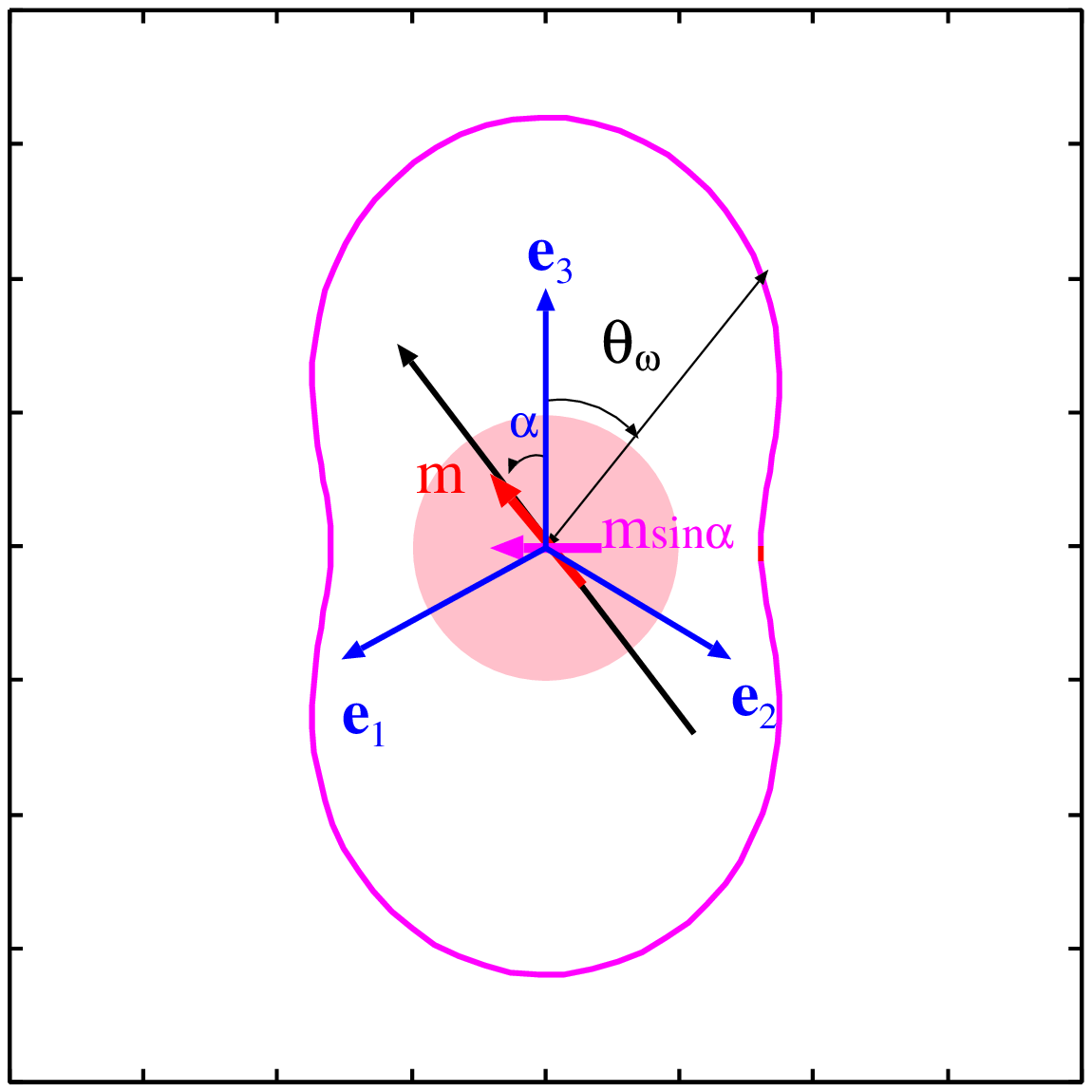}
\end{center}
\caption{Angular distribution of the luminosity.}
\label{emission}
\end{minipage}
\end{figure}

Although it is not well understood yet about where or how the pulsar emission takes place, 
the magnetic dipole model of pulsars simply states, from the viewpoint of energetics,  how the rotation energy of neutron stars is converted into pulsar emission \cite{SLS}. 
First of all, let us briefly see the physical mechanism. 
Consider the sphere with radius $a$ composed of uniform magnetization ${\bf M}$ (${\bf M}//{\hat z}$). Then the solution of the vector potential is
\beq
A_{\phi}=\frac{4\pi}{3}Ma^2\frac{r_<}{r_>}\sin\theta,
\eeq
where $(r_>, r_<)$ are the larger and smaller of $(r, R)$ \cite{jac}. The magnetic field ${\bf B}$ is then given by
\beqa
B_r&=&\frac{1}{r\sin\theta}\frac{\partial}{\partial\theta}(\sin\theta A_\phi)\nonumber\\
B_\theta&=&-\frac{1}{r}\frac{\partial}{\partial r}(rA_\phi) \nonumber\\
B_{\phi}&=& 0.
\eeqa
The magnetic field outside the sphere is 
\beqa
B^{({\rm out})}_\theta&=&\frac{|{\bf m}|\sin\theta}{r^3}\nonumber\\
B^{({\rm out})}_r&=&\frac{2|{\bf m}|\cos\theta}{r^3}.
\label{bout}
\eeqa
Hence it takes the maximum magnitude at $\theta=0$,
\beq
B^{({\rm out})}_{\rm max}=2|{\bf m}|r^{-3},
\label{maxb}
\eeq 
with the total magnetic moment ${\bf m}=4\pi/3a^3 {\bf M}$. 
When we consider a star with radius $R$ ($a=R$), $|{\bf m}|$ can be related to the magnetic field at the magnetic pole of the star, 
$B_p\equiv 2|{\bf m}|R^{-3}$, 
by way of Eq.~(\ref{maxb}). 
Accordingly the magnetic-field lines can be drawn by solving the differential equation,
\beq
\frac{dr}{B^{({\rm out})}_r}=\frac{rd\theta}{B^{({\rm out})}_\theta}
\eeq
to give 
\beq
r=R(\sin\theta/\sin\theta_0)^2,
\eeq
outside the star (Fig.~\ref{dipole}).

The magnetic field is uniform inside the star,
\beq
\bB^{({\rm in})}=\frac{2\bf m}{R^3}.
\label{bin}
\eeq

The {\it light cylinder} is defined as a cylinder with an axis along the pulsar rotation axis and with the radius $r_c=c/\omega$, where the velocity of the co-rotating frame with the pulsar attains the speed of light. Then the emission activity of the pulsar has been considered to originate from the magnetosphere with the radius $r_c$, while its mechanism is not clear yet \cite{SLS}. 

Denoting the inclined angle of the magnetic moment $\bf m$ to the rotation axis ${\bf e}_3$ by $\alpha$, we can evaluate the luminosity (energy radiation per unit time) $L_m$.
Then we can immediately see that the component perpendicular to ${\bf e}_3$, ${\bf m}_\perp$, is responsible to the luminosity, which is rotating in the plane perpendicular to the rotation axis, 
${\bf m}_\perp=|{\bf m}\sin\alpha|[{\bf e}_1\cos(\omega t)+{\bf e}_2\sin(\omega t)]$, with ${\bf e}_i\cdot{\bf e}_j=\delta_{ij}$. 
Thus the luminosity becomes equivalent with the one brought about by two oscillating magnetic dipoles (see Fig.~\ref{emission}).
A standard formula gives
\beq
\frac{d L_m}{d\Omega_\omega}=\frac{\omega^4}{4\pi c^3}|{\bf m}\sin\alpha|^2\frac{1}{2}(1+\cos^2\theta_\omega)
\eeq
for differential luminosity, where $\theta_\omega$ is an angle relative to the rotation axis \cite{jac}. 

Finally we have 
\beqa
L_m&=&\int d\Omega_\omega \frac{d L_m}{d\Omega_\omega}\nonumber\\
&=&\frac{2\omega^4}{3c^3}|{\bf m}\sin\alpha|^2\nonumber\\
&=&\frac{B_p^2R^6\omega^4\sin^2\alpha}{6c^3},
\label{lumi}
\eeqa 
for total luminosity.
It would be interesting here to see that the luminosity (\ref{lumi}) can be recast into the form,
\beq
L_m\simeq c\left(\frac{B_pR^3}{2r_c^3}\right)^2 r_c^2\sim c\frac{B(r_c)^2}{8\pi}\cdot 4\pi r_c^2,
\eeq
which implies that the energy density of the magnetic field ($\sim B(r_c)^2/8\pi$) flows outward through the surface of $4\pi r_c^2$ with the velocity of light $c$.

Assuming that this energy release originates from the rotational energy of neutron star,
\beq
\frac{d E_{\rm rot}}{dt}=-L_m,
\label{rote}
\eeq
where the rotation energy is given as  
$
E_{\rm rot}=\frac{1}{2}I\omega^2
$
with the moment of inertia $I$, assuming the rigid body. 
From Eqs,~(\ref{lumi}), (\ref{rote}), we have
\beq
P{\dot P}=\frac{2\pi^2 R^6\sin^2\alpha}{3Ic^3}B_p^2,
\eeq
where $P=2\pi/\omega$ is the period of the rotation.
The {\it characteristic age} $\tau_c$ is defined by 
\beq
\tau_c\equiv P/2{\dot P},
\label{age}
\eeq
which gives a rough estimate of the age of pulsars.

The spin-down rate $\dot \omega$ is given as ${\dot\omega}\propto -\omega^3$ from Eqs.~(\ref{lumi}) and (\ref{rote}), so that the {\it braking index} $n\equiv -\omega{\ddot\omega}/({\dot\omega})^2$ is three in this model. Note that the observed values of the braking index are less than three, e.g. $n=2.515\pm 0.005$ for the Crab, and it is usually considered that the significant fraction of the rotation energy is carried away by a pulsar wind \cite{bec}. Thus the magnetic dipole model is not sufficient to understand the emission processes in pulsars, while it can provide plausible estimates for the magnetic field and the characteristic ages.

Nowadays observation of $P-{\dot P}$ for pulsars are summarized on the plane \cite{bec}, where we can see three clusters of pulsars: ordinary radio pulsars, the millisecond pulsars and {\it magnetars}. We can easily see that most of the radio pulsars are centered around $10^{12}$G.  Magnetars are compact stars with huge magnetic field ($B=O(10^{14-15})$G), including soft $\gamma$-ray repeaters (SGRs) or anomalous X-ray pulsars (AXPs)\cite{woo,har,mere}. It exceeds the critical magnetic field 
defined as 
\beq
B_{\rm QED}=\frac{m_e^2c^3}{e\hbar}\simeq 4.4\times 10^{13}{\rm G},
\eeq  
which is obtained by equating the spacing of the Landau levels, $\delta E_e=\hbar eB/(m_e c)$, with the mass, $m_ec^2$. So it presents a criterion about how the relativistic effects are important and the quantum effects such as $e^+e^-$ creation or photon splitting becomes significant. It should be also interesting to recall the relation, $1{\rm MeV}^2\simeq 10^{13}$G in the natural unit. So, $B=10^{15}$G expected in magnetars may correspond to $10$MeV, which gets to the strong-interaction energy scale or the Fermi energy of nuclear matter or quark matter at nuclear density. 

We can easily asses how the magnetic field affects EOS inside compact stars. As shown in Eq.~(\ref{bin}) we can consider the uniform magnetic field inside the stars. Then $\delta E_p$ is small for protons, $\delta E_p\ll m_p$, while it is comparable with light quark mass, $\delta E_q\sim m_q$, for $B=O(10^{15})$G. On the other hand, the magnetic-interaction energy can be estimated by a simple formula, $E_{\rm mag}=\mu_i B$, with the Dirac magnetic moment, $\mu_i=e_i/(2m_i)$, for $i$-th particle having mass $m_i$ and electric charge $e_i$. For electrons it gives $O(\rm keV)$ for the canonical value of $B=O(10^{12})$G, which is comparable with the atomic-energy scale. So we may  easily expect that thermodynamic quantities or EOS of the neutron-star envelope or the crust region should be much affected by the magnetic field \cite{hae}.   
On the other hand, it becomes very tiny for protons ($E_{\rm mag}\sim {\rm keV}$) even for $B=O(10^{15})$G, 
which implies that the magnetic field little affect the EOS of nuclear matter.
Since it amounts to $E_{\rm mag}\sim{\rm MeV}$ for light quarks, the magnetic field looks to modify the EOS of quark matter. However, the Fermi energy should be very large ($O(100) {\rm MeV} \gg \delta E_q$), and a large number of the Landau levels are occupied, which may be well approximated by the usual Fermi sphere. Thus we can conclude the effect of the magnetic field little modifies the EOS in the core region, even for $B\sim 10^{15}$G.

Other high-field radio emitting pulsars with $B\ge 10^{14}$G or rotation-powered pulsar/magnetar transition objects have been also observed \cite{gav}. These observations may give a hint about the relationship between magnetic activity and neutron star spin-down.

The origin of such strong magnetic field has been a basic but a long-standing problem since the first discovery of pulsars, while many people believe that it originates from dynamo scenario due to the charged current or inheritance of the magnetic field from the progenitor main-sequence stars (fossil-field hypothesis) \cite{chan}. This problem becomes a current issue, stimulated by the discoveries of magnetars.
Due to the dynamo scenario, magnetic field is produced or maintained by the rotation or convection of charged fluid, but seems irrelevant for generation of high magnetic field in magnetars. It requires that magnetars be born with very short rotation periods of the order of 1-2 ms, and may not be supported by observations by two reasons \cite{mere}. First, The combination of high magnetic field and very rapid rotation is expected to impart a high velocity to the neutron star. However, up to now, the observational evidence for large spatial velocities in SGRs and AXPs is poor. Secondly, a large fraction of the rotational energy of a newly born magnetar, a few $10^{52}$erg, is lost due to the strong magnetic braking. However, an estimate of the explosion energy of the remnants containing magnetars yields values close to the canonical supernova explosion energy of $10^{51}$erg, implying initial periods longer than 5 ms.
 
Fossil-field hypothesis assumes the conservation of magnetic flux during the evolution of main-sequence progenitor star to a compact star: $B=(R_M/R)^2B_M$ with $R_M,B_M$ being the radius and magnetic field of the progenitor, respectively. It should be interesting to compare the radius $R$ with . Taking the sun as a typical main-sequence star, $(B_\odot)_{{\rm Max}}\simeq {\rm several~thousands}$~G and $R_\odot\simeq 7\times 10^{10}$cm. Squeezing the magnetic flux from $R_\odot$ to $R$, we have $R=R_\odot(B_\odot/B)^{1/2}\simeq 10$km for usual pulsars with $B=10^{12}$G, which should be consistent with standard neutron stars. However, $R<{\rm several}\times 10^{4-5}$cm for $B\simeq 10^{14-15}$G, which should be compared with the Schwartzschild radius, $R_S=2GM/c^2=3(M/M_\odot)\simeq 3\times 10^5$cm for $M \simeq M_\odot$. Thus we can see that $R$ should be comparable with $R_s$ and may fall below it in the extreme case, if the fossil-field hypothesis is applied to the generation of the magnetic field in magnetars. 

There is another possibility: a microscopic origin due to ferromagnetism or spin polarization of hadron matter \cite{nov}. Actually Makishima suggested the hadronic origin of the magnetic field in binary X-ray pulsars or radio pulsars \cite{mak}, since the accumulation of the observational data shows a peak with a narrow width.
Sometimes high-density nuclear matter is expected to show a resemblance with $^3$He system, by the scaling argument\cite{tam}, where quantum effects are essential and it shows the $^3P$-type superfluidity in some physical conditions.  Correspondingly neutron matter has been shown to exhibit the $^3P_2$ superfluidity around the nuclear density \cite{tam}. It is also well-known that $^3$He is very close to a ferromagnetic state under the high pressure due to the large magnetic susceptibility.
Microscopic calculations of nuclear matter have been repeatedly performed to 
find out a possibility of spontaneous spin polarization inside pulsars, but the negative results have been reported so far \cite{mat}. The magnetic susceptibility of nuclear matter increases monotonously with density, so that the ferromagnetic phase is not be expected in nuclear matter at any density. At high-density we can also expect quark matter and we may be tempted to ask the possibility in quark matter. Nowadays there have been actively studied about the possible evidences of hadron-quark transition in astrophysical phenomena as well as relativistic heavy-ion collisions \cite{bec,R-star2,sag}. If such phase transition occurs inside compact stars, it may affect thermal and magnetic properties as well as equation of state (EOS). Here we ask a microscopic origin of the magnetic field by considering uniform quark matter inside compact stars \cite{tat00}.

We consider the possibility of spontaneous spin polarization by the use of QCD. First, the evaluation of the total energy for spin-polarized quark matter has been done by using the one-gluon-exchange (OGE) interaction at zero temperature \cite{tat00,nie05,pal09}. Subsequently, the magnetic susceptibility has been studied within the Landau Fermi-liquid theory \cite{bay}, taking into account the screening effect for gluon propagation \cite{tat082,tat083,pal10}. Finally we present a magnetic phase diagram on the density-temperature plane. 

We also intend to address another interesting magnetic aspect of quark matter, which is characterized by the spatial modulation of the magnetic moment \cite{DCDW}. It is an inhomogeneous phase accompanying the chiral transition. In the standard scenario, the quark condensate, which is the order parameter of spontaneous symmetry breaking (SSB) of chiral symmetry, decreases due to the Pauli principle as density is increased. It is eventually vanished at some density-temperature point, which means the restoration of chiral symmetry. However, this may not be a unique scenario. Recall the FFLO state in the superconductivity; it has a spatially non-uniform order parameter and is considered to appear in the vicinity of the critical point, when two Fermi spheres with different spins have different sizes \cite{fflo}.  
Recent studies have shown that there possibly appear various kinds of inhomogeneous phases in the vicinity of the chiral-symmetry restoration \cite{bas,nic,brin}. Among them we consider a special one called dual-chiral-density-wave (DCDW) state in detail, because it exhibits many interesting theoretical features and is expected to bear various implications in the light of compact-star phenomena.  

Nowadays there have been done many theoretical studies about the deconfinement transition in high-density nuclear matter by using the MIT bag model \cite{mit} or other effective models of QCD to find the EOS including quark degrees of freedom. Unfortunately the lattice simulation is not possible, because the numerical calculation  suffers from the so called sign problem \cite{R-lat}. Thus there are ambiguities about the properties and the critical density of the deconfinement transition. 
Here we only assume the presence of quark matter without resource to EOS and the details of the deconfinement transition.

\section{Spontaneous spin polarization in quark matter}

A simple idea about spontaneous spin polarization owes to Bloch, who first discussed the emergence of ferromagnetism in electron gas at low density \cite{blo29,her66}. Consider the electron gas in the positively charged background to compensate the electromagnetic charge of electrons. Then the Coulomb interaction is classically vanished, but the Fock exchange interaction between electrons with the same spin gives an attractive effect due to the Pauli principle; in the non-relativistic approximation it reads  
\beq
E_{\rm ex}=-V\frac{9}{4}e^2\left(\frac{2}{9\pi}\right)^{1/3}n^{4/3}\left[(1+p)^{4/3}+(1-p)^{4/3}\right],
\eeq 
where $n$ is the total number density and $p$ is the polarization parameter defined by $p=(n_{s=+1}-n_{s=-1})/n$ with the number densities, $n_s$, of electrons with the spin $s=\pm 1$.
This spin dependence may be easily understood by observing that the repulsive Coulomb interaction is effectively avoided for electrons with the same spin, since they cannot approach due to the Pauli principle. On the other hand the kinetic energy is easily evaluated as
\beq
E_{\rm kin}=V\frac{\hbar^2}{2m}\frac{6}{5}\pi\left(\frac{9\pi}{2}\right)^{1/3}n^{5/3}\left[(1+p)^{5/3}+(1-p)^{5/3}\right],
\eeq
for non-relativistic electrons.
Note that the density dependence of each term is peculiar.  
Searching the minimum of the total energy $\varepsilon_{\rm total}(p)=(E_{\rm kin}+E_{\rm ex})/V$ with respect to $p$ by fixing the total number density $n_{+1}+n_{-1}=n$, we can observe that electron gas is completely polarized for $n^{1/3}<n_c^{1/3}\equiv \alpha/(1+2^{-1/3})$ with $\alpha\equiv 5/(6\pi^2)e^2m/\hbar^2(9\pi/2)^{1/3}$. The phase transition is of weakly first order in this case.

\begin{figure}[h]
\begin{minipage}{0.45\textwidth}
\begin{center}
\includegraphics[width=6cm]{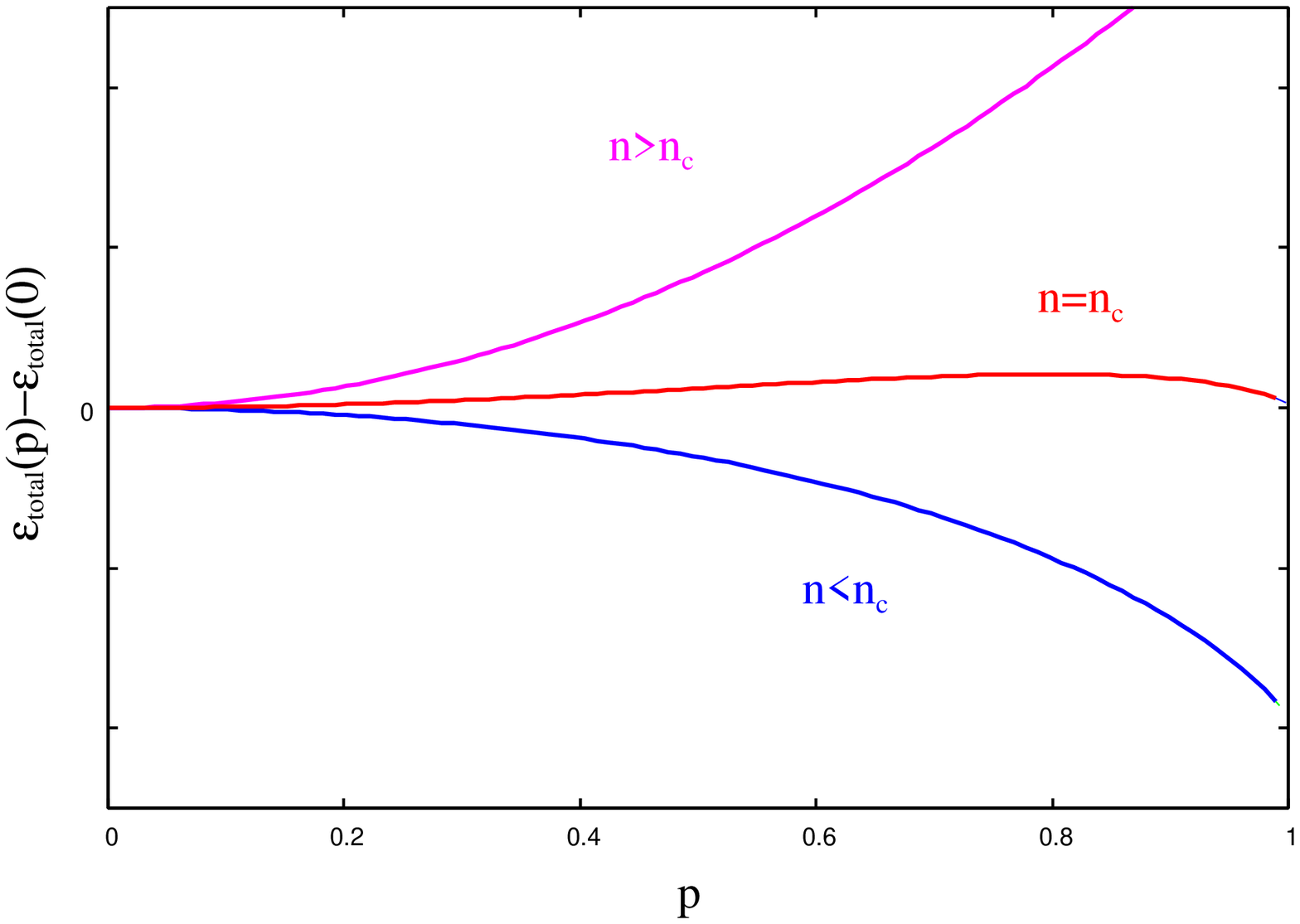}
\end{center}
\caption{Energy density (arbitrary scale) as the function of the polarization parameter $p=(n_{+1}-n_{-1})/n_q$.}
\label{bloch}
\end{minipage}
\hspace{\fill}
\begin{minipage}{0.54\textwidth}
\begin{center}
\includegraphics[width=8cm]{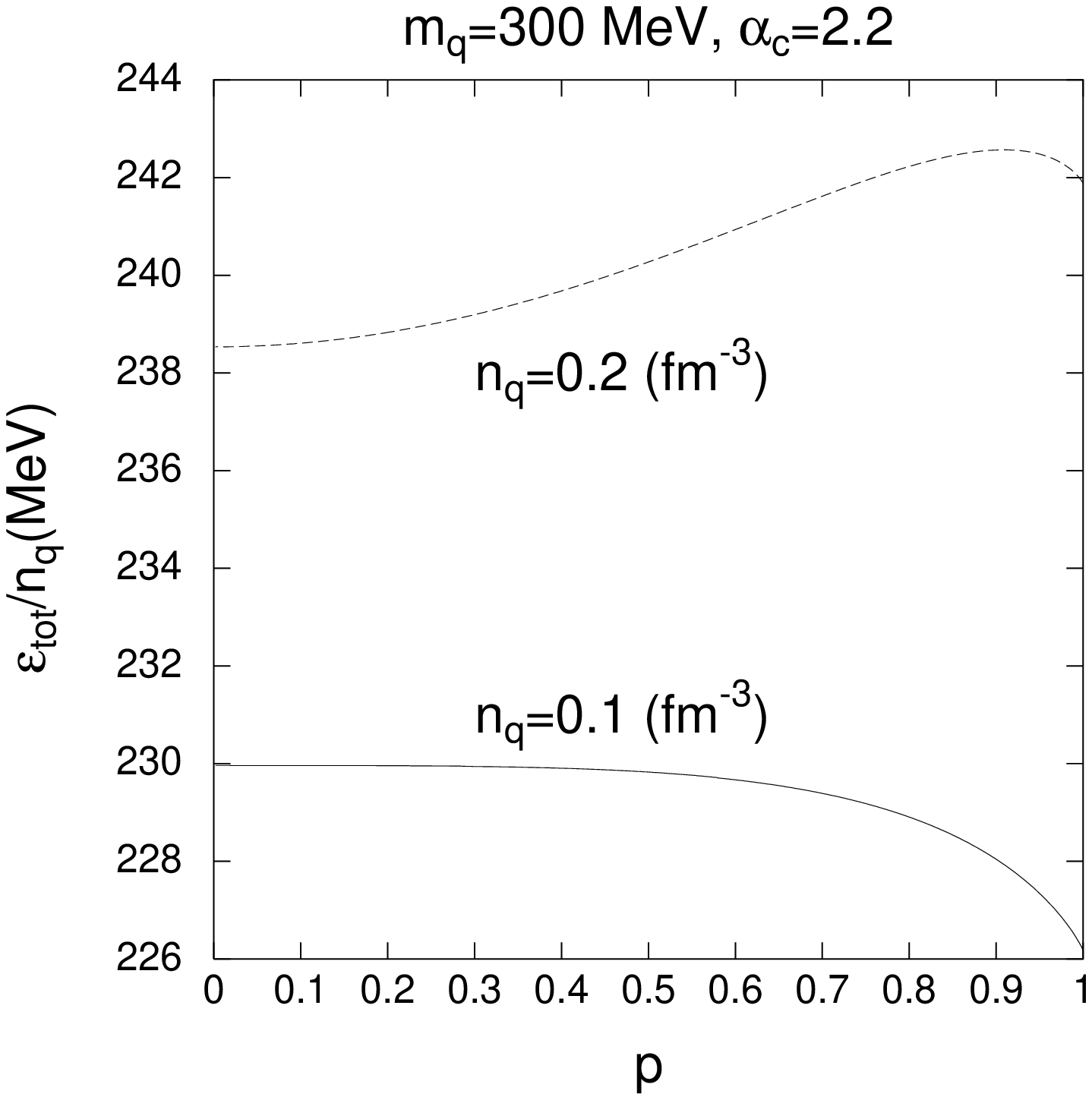}
\end{center}
\caption{Energy per particle for quark matter with the OGE interaction as a function of the polarization parameter, $p$.}
\label{ferro}
\end{minipage}
\end{figure}

This is the result within the Hartree-Fock approximation, but it has been also shown 
by the quantum Monte Carlo simulation that the electron gas is in the ferromagnetic phase at very low density \cite{ele}. 
Recently it has been experimentally observed \cite{you}. A lesson we learned here is that we need no spin dependent interaction in the original Lagrangian to see spontaneous spin polarization, but a symmetry principle plays an important role in this context.

When we consider quark matter within QCD, we notice that the situation seems to be similar to the electron gas; quark matter is color neutral as a whole, so that the exchange term of the one-gluon-exchange (OGE) interaction gives a leading-order contribution to the total energy. In ref. we have calculated the interaction energy of relativistic quarks in a perturbative way, and demonstrated the spontaneous spin polarization around nuclear density, $n_q\simeq \rho_0\simeq 0.16$fm$^{-3}$. 

Assuming that spin-polarized quark matter with density $n_q$ inside compact stars, we can roughly estimate the magnetic field at the surface (see Eq.~\ref{maxb})),  
\begin{eqnarray}
B_{\rm max}&=&\frac{8\pi}{3}\left(\frac{r_q}{R}\right)^3\mu_qn_q\nonumber\\
&\simeq&10^{15}{\rm G}\left(\frac{r_q}{R}\right)^3\left(\frac{\mu_q}{\mu_N}\right)\left(\frac{n_q}{0.1{\rm fm}^{-3}}\right)
\end{eqnarray}  
for the quark core of the radius $r_q$, where $\mu_q$ is the quark magnetic moment and $\mu_N$ the nuclear magneton
\footnote{One may wonder about the back-reaction of the generated magnetic field to quarks. The Fermi energy is $O(100)$MeV and the quark mass is rather heavy in this case, so that the Landau orbiting and the magnetic interaction are safely discarded.}
.

\subsection{Magnetic susceptibility within the Landau Fermi-liquid theory}

In the recent papers we have studied the magnetic susceptibility within the Landau Fermi-liquid theory \cite{tat082,tat083}, to get more insight about the phase transition and derive a phase diagram in the density-temperature plane.  

 Applying a tiny and uniform magnetic field $\bB$ (along the $z$-axis), we can study the linear response of quark matter by calculating the change of magnetization, $\bf M$, by the external field. Then the magnetic susceptibility $\chi_M$ is defined as $\chi_M=\partial {\bf M}/\partial \bB|_{\bB=0}$.
We can easily see that $\chi_M$ measures the spin-spin correlation in the normal quark matter, or the curvature of the free energy at the origin with respect to the order parameter, the magnetization $\bf M$.
The free energy is given as a function of the external magnetic field, $F(\bB)$, from which $\bf M$ is given by 
\beq
{\bf M}=-\frac{\partial F(\bB)}{\partial \bB}.
\eeq
Defining the Legendre transform of $F$ such that 
\beq
G({\bf M})=F+{\bf M}\cdot \bB,
\eeq
$G$ satisfies the reciprocity relation,
\beq
\frac{\partial G(\bf M)}{\partial{\bf M}}=\bB.
\eeq
Therefore $\chi_M$ can be written as 
\beq
\chi_M=\left[\frac{\partial^2 G(\bf M)}{\partial{\bf M}^2}\right]^{-1},
\eeq
which is the inverse of the curvature of $G$ at the origin, assuming ${\bf M}=0$ for $\bB=0$.
\begin{figure}[h]
\begin{center}
\includegraphics[width=0.6\textwidth]{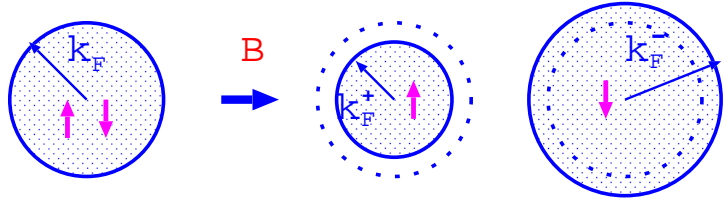}
\caption{Two Fermi spheres with different spins have different sizes to generate the magnetization 
after applying an external magnetic field.}
\label{flt}
\end{center}
\end{figure}
Adding $-{\bf M}\cdot\bB$ to the free energy $G({\bf M})$ we find
the fluctuation amplitude at $T$,
\beq
\langle M^2\rangle|_{\bB=0}=T\chi_M,
\label{fluc1}
\eeq
which means the fluctuation amplitude diverges as $\chi_M$ \cite{and}.  

$\chi_M$ can be written in terms of the Landau-Migdal parameters derived from the quark-quark interaction,
\beq
\chi_M=\left(\frac{{\bar g}_D\mu_q}{2}\right)^2\frac{N(T)}{1+N(T)\bar
f^a},
\label{chim}
\eeq
where ${\bar g}_D\equiv \int_{|{\bf k}|=k_F} d \Omega_{{\bf k}}/4\pi g_D({\bf k})$ is the effective gyromagnetic ratio. $N(T)$  is the effective density of states around the Fermi surface, which is written as 
\beq
N(0)=\frac{N_ck_F^2}{\pi^2 v_F}
\eeq 
at $T=0$, where the Fermi velocity $v_F$ is given by the spin-independent Landau-Migdal parameter $f_1^s$, $k_F/E_F-(N_ck_F^2/3\pi^2)f_1^s$. $\bar f^a$ is the spin-dependent Landau-Migdal parameter. The divergence of $\chi_M$ or $1+N(T)\bar f^a=0$ implies the phase transition to the ferromagnetic phase.

\begin{figure}[h]
\begin{center}
\includegraphics[width=0.4\textwidth]{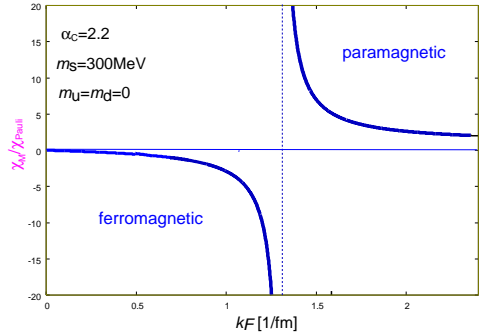}
\caption{Spin susceptibility $\chi_M$ at $T=0$ within the Landau's Fermi liquid theory. $\chi_{\rm Pauli}$ is the Pauli paramagnetism. It diverges around $k_F\simeq 1.4$fm$^{-3}$, the order of nuclear density, $\rho_0$, and quark matter is in the ferromagnetic phase at lower density side.}
\end{center}
\label{suscep}
\end{figure}

We present here a heuristic argument by the use of the naive OGE interaction.
\beq
{\mathcal L}_{\rm QCD}=-\frac{1}{2}G^{\mu\nu}G_{\mu\nu}+\sum_f {\bar q}_f\left(i\gamma^\mu D_\mu-m_f\right)q_f,
\eeq
where 
\beqa
G_{\mu\nu}&=&\partial_\mu A_\nu-\partial_\nu A_\mu-ig[A_\mu, A_\nu]\nonumber\\
D_\mu q_f&\equiv&(\partial_\mu-igA_\mu) q_f
\eeqa
with the gluon field, $A_\mu=\sum_{a=1}^8 A_\mu^a\lambda^a/2$.
Since spin is coupled with motion in relativistic theories, we must define spin polarization in a proper way. A relevant one is to introduce the space-like four vector $a^\mu$ with the constraints, $a\cdot k=0$ and $a^2=-1$ for a particle with momentum $\vk$. The standard choice may be 
\beq
\va=\bzeta+\frac{\vk(\bzeta\cdot\vk)}{m(E_k+m)},~~~
a^0=\frac{\vk\cdot\bzeta}{m},
\label{spinvec}
\eeq
 where the three vector $\bzeta$ specifies the direction of spin in the rest frame of each particle \cite{ber}. Then we have the polarization density matrix $\rho(k,\zeta)$,
\beq
\rho(k,\zeta)=\frac{1}{2m}(\ds{k}+m)P(a),
\eeq
with the projection operator, $P(a)=(1+\gamma_5\ds{a})/2$. Given the quasiparticle interaction $f_{\vk\zeta a;\vq\zeta' b}$ between quarks with momentum $\vk$, spin polarization $\zeta$ and color $a$ and $\vq$, $\zeta'$ and $b$, the color symmetric interaction is generally given as
\beq
f_{\bk\zeta,\bq\zeta^{\prime}}
=\frac{1}{N_c^2}\sum_{a,b}f_{\bk\zeta a ,\bq\zeta^\prime b }
=\frac{m}{E_k}\frac{m}{E_q}M_{\bk\zeta,\bq\zeta^{\prime}},
\eeq
with the invariant matrix element within the OGE interaction,
\beq
M_{\bk\zeta,\bq\zeta^{\prime}}=-g^2\frac{1}{N_c^2}{\rm
tr}\left(\lambda_\alpha/2\lambda_\alpha/2\right)M^{\mu\nu}(k,\zeta; q,\zeta')D_{\mu\nu}(k-q).
\label{invmatrix}
\eeq
Here $D_{\mu\nu}$ is the gluon propagator and $M^{\mu\nu}(k,\zeta; q,\zeta')$ the interaction tensor,
\beq
M^{\mu\nu}(k,\zeta; q,\zeta')={\rm
tr}\left[\gamma^\mu\rho(k,\zeta)\gamma^\nu\rho(q,\zeta')\right],
\label{inttensor}
\eeq
Taking the Feynman gauge for the gluon field, the invariant matrix element can be explicitly written as
\beq
M_{\vk\zeta,\vq\zeta^{\prime}}=g^2\frac{N_c^2-1}{4N_c^2m^2}\left[2m^2-k\cdot q-m^2a\cdot b\right]\frac{1}{(k-q)^2}.
\eeq
In the non-relativistic limit, it is reduced to 
\beq
M_{\vk\zeta,\vq\zeta^{\prime}}\rightarrow
-g^2\frac{N_c^2-1}{4N_c^2}\frac{1+\bzeta\cdot\bzeta'}{(\vk-\vq)^2}<0. 
\eeq
Thus a pair with the parallel spin $(\bzeta=\bzeta')$ gives an attractive interaction in the nonrelativistic limit. 
This makes a base of the Bloch mechanism \cite{her66}: the Fock exchange interaction give rise to an attractive effect for the pair of the parallel spin due to the Pauli principle. 
Then the quasiparticle interaction on the Fermi surface is given as 
\beq
\left.f_{\vk\zeta,\vq\zeta^{\prime}}\right|_{|\vk|=|\vq|=k_F}=g^2
\frac{N_c^2-1}{8N_c^2 E_F^2}
\left[2m^2-E_F^2+k_F^2\cos\theta_{\widehat{\vk\vq}}-m^2 a \cdot b\right]
\frac{1}{-k_F^2(1-\cos\theta_{{\widehat{\vk\vq}}})}. 
\eeq  
We can immediately see that it diverges for the collinear momenta,
$\vk=\vq$. 
Substituting the explicit formula of the spin vector in Eq.~(\ref{spinvec}), we can see that it consists of two parts, the spin-independent ($f^s_{\bk,\bq}$) and 
spin-independent ($f^a_{\bk,\bq}$) terms;
\beq
f_{\bk\zeta,\bq\zeta'}=f^s_{\bk,\bq}+\zeta\zeta'f^a_{\bk,\bq},
\label{qpint}
\eeq 
from which we can derive the Landau-Migdal parameters by the angle-integral over the Fermi surface.
For the spin-independent Landau-Migdal parameter, $f_1^s$, 
\beqa 
f_1^s=-3\frac{g^2(N_c^2-1)}{8N_c^2 E_F^2}\frac{m^2}{2k_F^2}
\int_{-1}^1d(\cos\theta)\frac{\cos\theta}{1-\cos\theta}\rightarrow \infty,
\label{f1s0}
\eeqa
the spin-dependent one, $\bar f^a$,
\beq
{\bar f^a}\equiv \int\frac{d\Omega_{\vk}}{4\pi}\int\frac{d\Omega_{\bq}}
{4\pi}\left.f^a_{\bk,\bq}\right|_{|\bk|=|\bq|=k_F}=-\frac{g^2(N_c^2-1)}{8N_c^2E_F^2}
\frac{m(2E_F+m)}{3k_F^2}+f_1^s/3.  
\label{fabar0}
\eeq
Putting Eqs.~(\ref{f1s0}) and (\ref{fabar0}) in Eq.~(\ref{chim}), we have the final expression, 
\beq
\left(\chi_M/ \chi_{\rm Pauli}\right)^{-1}=1
-\frac{C_f g^2}{12\pi^2E_Fk_F}m(2E_F+m)
\label{wosc}
\eeq
with the Casimir operator per color, 
$C_f=\frac{N_c^2-1}{2N_c}$, where $\chi_{\rm Pauli}$ is the susceptibility of non-interacting Fermi gas, 
$
\chi_{\rm Pauli}={\bar g}_D^{2}\mu_q^2 N_c k_FE_F/4\pi^2.
$
. Thus we can observe that infrared divergences cancel each other in the expression of $\chi_M$. It is to be noted that 
$\chi_M$ is proportional to the quark mass. Thus, we may say that 
heavier quarks favor the spontaneous magnetization. Taking the non-relativistic limit, $m\gg k_F$, and replacing $C_f$ and $g$ by one and $e$, respectively, we can recover the standard formula for electron gas interacting with the Coulomb potential\cite{her66},
\beq
\left(\chi/\chi_0\right)^{-1}=1-\frac{e^2 m}{k_F\pi}
\eeq
In Fig.~\ref{suscep} we present a result by using the MIT bag model parameters.

\subsection{Magnetic phase diagram}

We have also studied the screening effect for OGE interaction on $\chi_M$, since it is well known that such many-body effect is important in the electron gas. Actually the screening effect disfavors the spin alignment \cite{bru57}. 
Moreover, it is necessary to include it when we consider the gauge interaction like QED/QCD, since the infrared behaviour is singular and it is improved by the screening effect.
In QCD it has another interesting effect, depending on the number of flavors: it may favor the ferromagnetic transition for a large number of flavors. In the following let us briefly see the screening effect.

The gluon propagator $D_{\mu\nu}(p)$ is generally written as 
\beq
D_{\mu\nu}(p)=P^t_{\mu\nu}D_t(p)+P^l_{\mu\nu}D_l(p)-\xi\frac{p_\mu
p_\nu}{p^4}, 
\eeq
where 
$P^{t(l)}_{\mu\nu}$ is the projection operator onto the
transverse (longitudinal) mode, 
\beqa
P^t_{\mu\nu}&=&(1-g_{\mu 0})(1-g_{\nu
0})\left(-g_{\mu\nu}-\frac{p_\mu p_\nu}{|\bp|^2}\right)\nonumber\\
P^l_{\mu\nu}&=&-g_{\mu\nu}+\frac{p_\mu
p_\nu}{p^2}-P^t_{\mu\nu}.
\eeqa
Correspondingly, $D_{t(l)}(p)$ are the propagator of transverse (longitudinal) gluons, 
modified by the medium effect, which can be described by the Debye screening \cite{LeBe}.
The self-energies for the transverse and longitudinal gluons are calculated by the hard dense loop (HDL) 
resummation to give 
\beqa
\Pi_l(p_0,{\bf p})&=&\sum_{f=u,d,s} \left( m_{D,f}^2+i\frac{\pi
m_{D,f}^2}{2u_{F,f}}\frac{p_0}{|\bp|} \right)\nonumber\\
\Pi_t(p_0,{\bf p})&=&-i\sum_{f=u,d,s} \frac{\pi u_{F,f} m_{D,f}^2}{4}\frac{p_0}{|\bp|}, 
\eeqa
in the limit $p_0/|\bp|\rightarrow 0$, with the Fermi velocity $u_{F,f}\equiv
k_{F,f}/E_{F,f}$ and the Debye mass,
$m_{D,f}^2\equiv g^2\mu_f k_{F,f}/2\pi^2$ for each flavor $f$ \cite{LeBe}
\footnote{The Debye mass is given as $e^2\mu^2u_F/\pi^2$ for 
electron gas in QED.}
. 
The appearance of the imaginary part in $\Pi_t$ 
physically means the damping of the transverse gluons due to the interaction with the surrounding quarks (Landau damping).
Thus the longitudinal gluons are statically
screened by the Debye mass, while the transverse gluons are dynamically screened due to the Landau damping. 
Accordingly, the screening effect for the transverse gluons is ineffective 
at $T=0$, where soft gluons ($p_0/|\bp|\rightarrow 0$) contribute. 
At finite temperature, gluons with $p_0 \sim O(T)$ can contribute due to the diffuseness of the Fermi surface and the transverse gluons
are effectively screened, which gives rise to another interesting consequence for $\chi_M$. 
We have seen a non-Fermi liquid behaviour, which is inherent in the gauge theories: there appears $T^2\ln T$ term in $\chi_M$, 
besides the usual $T^2$ term \cite{tat082,tat083}. 

Taking into account the screening effect, we can evaluate the magnetic susceptibility. 
To summarize, we present in Fig.~\ref{Fig:diagram} the phase diagram of ferromagnetic quark matter on the temperature-density plane. 

\begin{figure}[h]
\begin{center}
\includegraphics[width=0.4\textwidth]{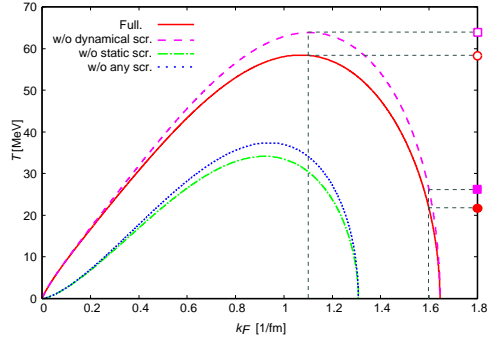}
\caption{Magnetic phase diagram in the density-temperature plane. The open (filled) circle indicates 
the Curie temperature at $k_F=1.1(1.6)$ fm$^{-1}$ while the squares show those without the $T^2 \ln T$ term.}
\label{Fig:diagram}
\end{center}
\end{figure}

\subsection{Spin wave}

Ferromagnetic quark matter is specified by the non-vanishing magnetization ${\bf M}$, so that rotation symmetry is violated there, $SO(3)\rightarrow O(2)$. 
Such SSB should accompany the Nambu-Goldstone mode, {\it spin wave}. Different from the usual treatment of the spin wave in the Heisenberg model \cite{alt}, we must carefully formulate it since quarks freely move there. For the ferromagnetism of itinerant electrons, Herring took an intuitive but correct approach called the {\it spiral} approach \cite{her66,her52}. He introduced the collective variables as variants from the spin direction. 
In ref.\cite{exot} we have proposed a formulation within the path integral of the spin coherent-state. 
Since the spin wave or its quantized magnons plays important roles for the thermal evolution as well as the magnetic evolution of compact stars, more elaborate studies are needed about its properties including the dipersion relation: specific heat and thermal conductivity may be affected by the magnons and a new cooling process may open through  the absorption and emission of magnons in the interaction vertices.
Moreover, the magnon-exchange interaction may work in the ferromagnetic phase, which gives, e.g., an exotic superconductivity \cite{kar}.

\section{Chiral transition and Inhomogeneous phases}


Here we consider another type of magnetism in quark matter. Recently there are many works about the inhomogeneous phases accompanying the chiral transition \cite{bas,nic,brin}. 
According to the usual discussion about the chiral transition, the scalar condensate $\langle{\bar q}q\rangle$ is decreased as density or temperature is increased, and eventually vanished at some density \cite{R-star2}. Since the dynamical mass is proportional to the condensate, it also decreases. 

We, hereafter, consider two-flavor ($u,d$) quark matter for simplicity.
Chiral symmetry is then represented by $SU(2)_L\times SU(2)_R$ algebra, under which the quark field is transformed s.t.
\beqa
\psi\rightarrow \psi'&=&\psi+i\alpha^a(\tau^a/2)\psi\nonumber\\
\psi\rightarrow \psi'&=&\psi+i\beta^a(\tau^a/2)\gamma_5 \psi,
\eeqa  
for small parameters, $\alpha^a,\beta^a\ll 1$. The first one is the usual isospin transformation, while the second one, called chiral transformation, mixes the different parity states. Many studies have shown that Lagrangian should be chirally invariant for massless quarks, but it is spontaneously broken in the vacuum, which is specified by the non-vanishing $q\bar q$ condensate, $\langle{\bar \psi}\psi\rangle\neq 0$. As important consequences, the quark mass can be dynamically generated by the spontaneous symmetry breaking (SSB) and pions emerge as a collective state of quarks and antiquarks. At finite density the Fermi sea prevent the formation of the condensate due to the Pauli principle, and chiral symmetry should be eventually restored at some density. 


Recent studies have suggested that the homogeneous ground state becomes unstable for producing the inhomogeneous phases in prior to the chiral transition. Here we consider one of such instabilities in detail by using the effective model of QCD, Nambu-Jona Lasinio (NJL) model \cite{DCDW,kut}. In the paper \cite{DCDW} we have demonstrated formation of a density wave in quark matter in the chiral limit: where not only scalar density but also pseudoscalar density takes a non-vanishing value,
\beqa
\langle{\bar \psi}\psi\rangle&=&\Delta\cos(\vq\cdot\br)\nonumber\\
\langle{\bar \psi}i\gamma_5\tau_3\psi\rangle&=&\Delta\sin(\vq\cdot\br).
\label{meanf}
\eeqa  
Note that we can utilize the pseudoscalar degree of freedom in quark matter, different from the vacuum, which must be an eigenstate of parity. Consequently the above configuration means a violation of parity. It should be interesting here to see a similarity with the FFLO state in superconductivity \cite{fflo}: a combination of both densities can be written as 
$\langle{\bar \psi}\psi\rangle+i\langle{\bar \psi}i\gamma_5\tau_3\psi\rangle=\Delta{\rm exp}(i\vq\cdot\br)$, which corresponds to the complex energy-gap parameter due to the condensation of the Cooper pairs \cite{schr}. In the FFLO state, the gap function, $\Delta({\br})\equiv \langle\psi_\downarrow({\br})\psi_\uparrow({\br})\rangle$, modulates in space with the wave number characterized by the center-of-mass momentum of the Cooper pairs. The similarity indicates a general mapping between superconductivity and magnetism \cite{scho}. It might be also interesting to see the phase transition in terms of the susceptibility. Similarly to the magnetic susceptibility in the previous section, we can introduce the susceptibility or the density correlation function in scalar or pseudoscalar channel, $\chi_{\rm s(\rm ps)}(\omega, q)$ \cite{DCDW}. In the ferromagnetic case, $\chi_M(\omega=0, q=0)$ diverges at the critical point, but $\chi_{\rm s(\rm ps)}(\omega=0, q\neq 0)$ in this case. The finite value of $q$ is  characteristic in the inhomogeneous phase.

\begin{figure}[h]
\begin{minipage}{0.4\textwidth}
\begin{center}
\includegraphics[width=4.8cm]{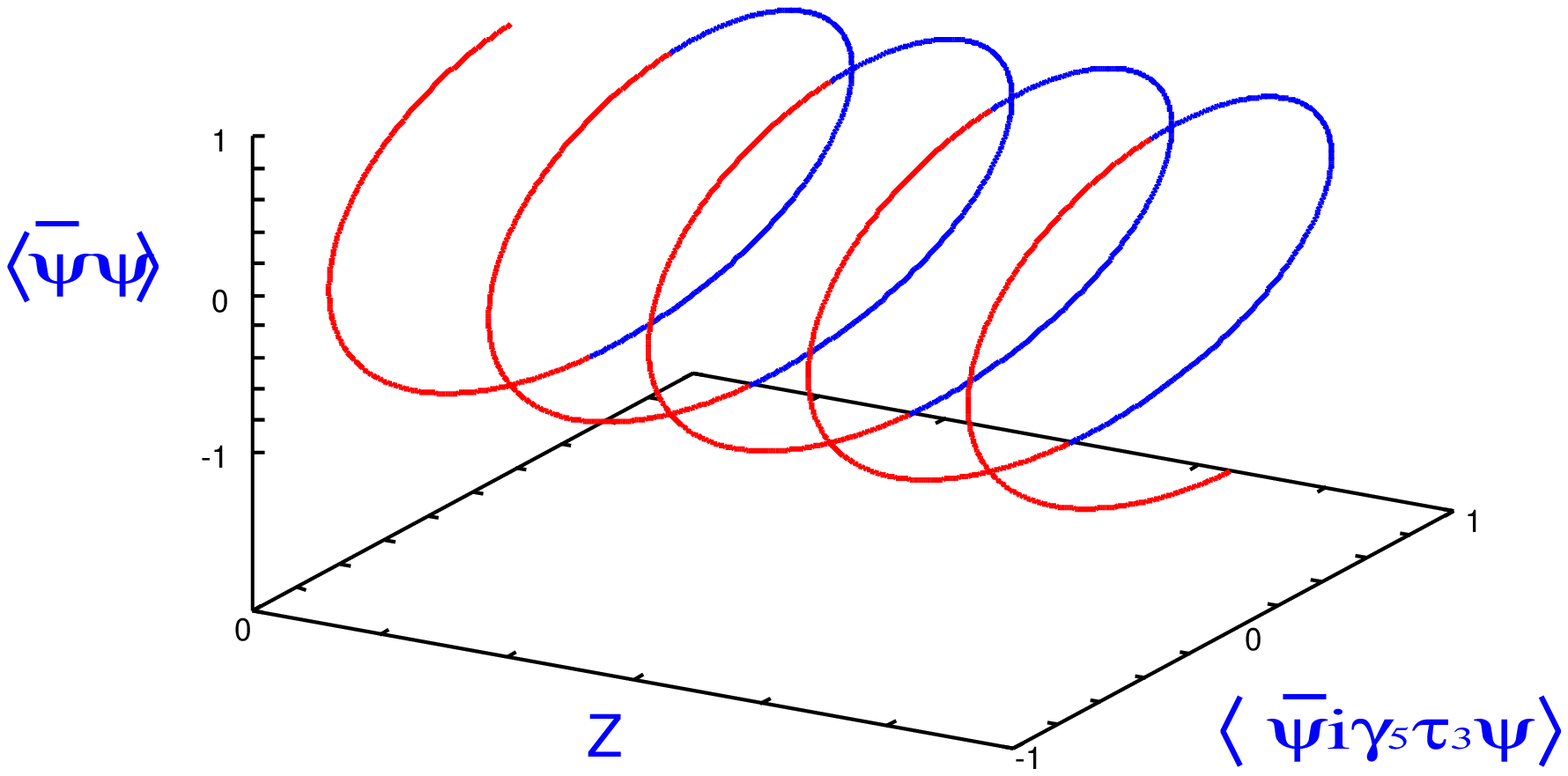}
\end{center}
\caption{Dual chiral density wave in the chiral space.}
\label{Fig:diagram}
\end{minipage}
\hspace{\fill}
\begin{minipage}{0.56\textwidth}
\begin{center}
\includegraphics[width=4.5cm]{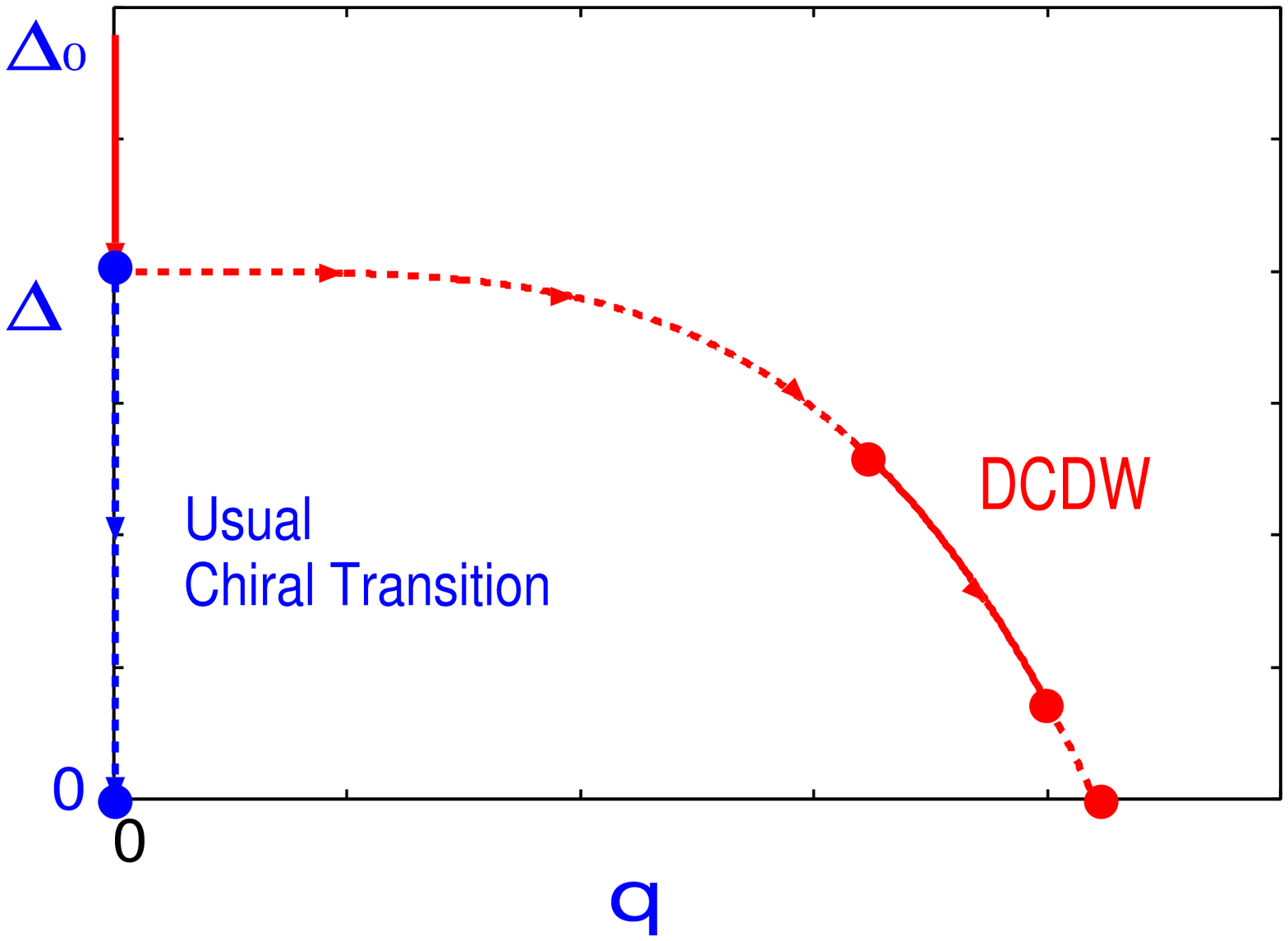}
\end{center}
\caption{Possible chiral restoration paths on the $\Delta-q$ plane. Dotted lines show the first order phase transitions.}
\label{path}
\end{minipage}
\end{figure}

Consider the 2-flavor Nambu-Jona Lasinio model as an effective model of QCD at low density,
\beq
{\mathcal L}_{\rm NJL}={\bar\psi}(i\ds{\partial}-m_c)\psi+G\left[\left({\bar\psi}\psi\right)^2+\left({\bar\psi}i\gamma_5\btau\psi\right)^2\right],
\eeq
where $G$ is the coupling constant and $m_c$ the current mass. 
\footnote{We, hereafter, consider the chiral limit where $m_c=0$ in this subsection.}
Putting Eq.~(\ref{meanf}) in the Lagrangian under the mean-field approximation, we have 
\beq
{\mathcal L}_{\rm MF}={\bar\psi}\left[i\ds{\partial}-M^*\left(\cos(\vq\cdot\vr)+i\gamma_5\tau_3\sin(\vq\cdot\vr)\right)\right]-\frac{M^{*2}}{4G},
\eeq
where the effective mass $M^*$ is defined by way of $M^*=-2G\Delta$. 
Introducing a new spinor by using a local chiral transformation,
\beq
\psi(\vr) = e^{-i\tau_3 \gamma_5 \vq\cdot \vr/2}\psi_w(\vr),
\label{chitrans}
\eeq
$\mathcal L_{\rm MF}$ is recast into
\beq
\mathcal{L}_{\rm MF}= \bar{\psi}_w \left[ i\ds{\partial} -M^* 
-\tau_3 \gamma_5\bgamma\cdot \vq/2 \right] \psi_w -\frac{M^{*2}}{4G}.
\eeq
This is a kind of the Weinberg transformation \cite{wein}, which makes the Lagrangian in the presence of inhomogeneous DCDW to the one with the homogeneous axial-vector field. 
\begin{table}[h]
\begin{center}  
\caption{Separation of amplitude and phase degrees of freedom by way of the Weinberg transformation.}
\vskip 0.5cm
\begin{tabular}{ccc}\hline
non-uniform&&uniform\\ \hline
&&\\
$\langle{\bar\psi\psi}\rangle=\Delta\cos\vq\cdot\vr$&$\rightarrow$&$\langle{\bar\psi}_w\psi_w\rangle=\Delta$\\ 
$\langle{\bar\psi}i\gamma_5\tau_3\psi\rangle=\Delta\sin\vq\cdot\vr$&$\rightarrow$&$\langle{\bar\psi}_wi\gamma_5\tau_3\psi_w\rangle=0$\\ 
---&$\rightarrow$&$\gamma_5\tau_3\bgamma\vq/2$\\ 
&&\\ \hline
\end{tabular}
\end{center}
\end{table}

Then the quark wave function is still given by the plane-wave solution $\psi_k(\vr)$ with the energy eigenvalue,
\beq
E^\pm (\vp) =\sqrt{ \vp^2+M^{*2} + \vq^2/4 \pm
\sqrt{(\vp\cdot \vq)^2 +M^{*2} \vq^2} }. 
\label{single}
\eeq 
Accordingly the Fermi seas are deformed in oblate and prolate shapes, depending on the spin degree of freedom. In Fig.~ we show energy spectra for example.
\begin{figure}[h]
\begin{center}
\includegraphics[width=0.4\textwidth]{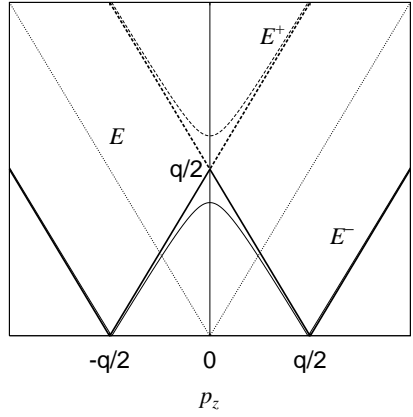}
\caption{Energy spectra for $p_\perp=0$ under the condition, $q/2>M^*$. $E$ is the one in the absence of DCDW.}
\end{center}
\end{figure}

Thermodynamic potential is given as a function of $M^*$ and $q$ at finite temperature $T$ and chemical potential $\mu$ :
\beqa
\Omega(q,M^*)&=&-T\log Z/V \nn
&=& -N_f N_c \int\frac{d^3p}{(2\pi)^3} \sum_{s=\pm} \left\{ T\ln \left[ e^{-\beta 
(E^s-\mu)}+1 \right]
\left[ e^{-\beta(E^s+\mu)}+1 \right] +E^s\right\} +\frac{M^{*2}}{4G}. 
\eeqa
The last term in the curly bracket denotes the vacuum contribution, 
which has an essential role in the chiral transition. 
Since single-particle spectra are deformed in momentum space, 
the simple momentum cut-off is irrelevant and sometimes leads to unphysical results. We employ the proper-time regularization with cutoff $\Lambda$ to evaluate the vacuum contribution.

\begin{figure}[h]
\begin{center}
\includegraphics[width=0.4\textwidth]{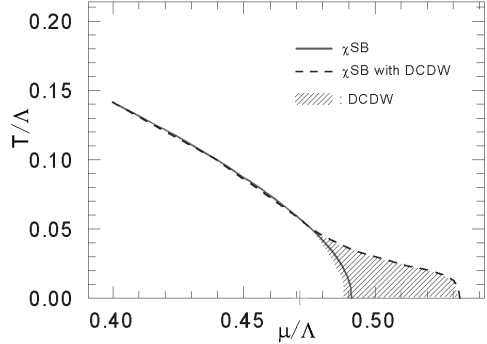}
\caption{Phase diagram of DCDW on the density-temperature plane. DCDW emerges in the vicinity of the chiral transition.}
\end{center}
\end{figure}

\subsection{"Nesting" mechanism}

\begin{figure}[h]
\begin{center}
\includegraphics[width=0.45\textwidth]{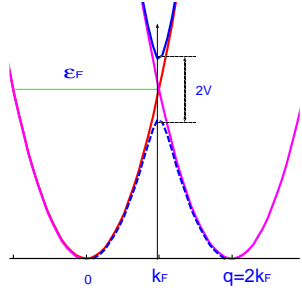}
\caption{Energy spectra of one dimensional electron gas in the presence of SDW.}
\label{nest}
\end{center}
\end{figure}

First we consider the one dimensional case. Assume the presence of the spin density wave (SDW) ${\rm exp}(\pm iqz)$ coupled with the non-relativistic 
electrons like 
\beq
H_{\rm int}=-V\left(e^{-iqz}u^\dagger\sigma_+u+e^{iqz}u^\dagger\sigma_-u\right).\eeq
Then eigenvalues are
\beq
E_k=\frac{1}{2}(\epsilon_k+\epsilon_{k+q})\pm\left[\frac{1}{2}(\epsilon_k-\epsilon_{k+q})^2+4V^2\right]^{1/2},
\eeq
with the free particle energy, $\epsilon_k=k^2/2m$ (Fig.~\ref{nest}). In the case of $q=2k_F$, there is produced a gap at the Fermi surface of the free electrons. Thereby the total energy is always decreased by the interaction, {\it independent} of the strength of the coupling. This is called a nesting effect of the Fermi surface \cite{ove,gru}. We can also see this by calculating the correlation function; it is given by the Lindhard function $L(\omega=0, q)$, which exhibits a logarithmic divergence at $q=2k_F$. In the higher dimensional cases, the nesting is incomplete but we can see its reminiscence. 

We can see the similar mechanism should be also responsible to the emergence of DCDW, but in somewhat different manner. By putting $\vp_\perp=0$, we can consider quasi-one dimensional case. The energy spectra exhibit different features, depending on the conditions, $q/2>M^*$ and $q/2<M^*$. It is not obvious which condition holds in the DCDW phase, but the numerical results shows that the former condition always holds. Then we can understand the mechanism of the DCDW formation by two steps. First looking at Fig.~\ref{nest}, we can see that the levels of massless quarks cross each other at $p_z=0$. After switching on the mass term the level crossing is finally disentangled to give $E^{\pm}$. So, if we set $q=2\mu$, there is always the energy gain, irrespective of the dynamics. 

\subsection{Some implications}

The presence of DCDW has some implications on compact star phenomena. Let's consider, for example, the $\beta$-decay of quarks in the DCDW state,
\beq
d (p_1)\rightarrow u(p_2)+e^-(p_3)+{\bar \nu}_e(p_4).
\label{betad}
\eeq 
From Eq.~(\ref{chitrans}) the DCDW state can be represented as a chirally rotated state,
\beq
|\psi_w\rangle={\hat U}({\bf q})|\psi\rangle,
\eeq
with 
\beq
{\hat U(q)}={\rm exp}(i\int A_3^0 {\bf q}\cdot{\bf r}d^3x),
\eeq
the quark current $h_{1+i2}^\mu\equiv {\bar \psi}\gamma^\mu(1-\gamma_5)\tau_+\psi$ is transformed as
\beq
{\tilde h}^\mu_{1+i2}\equiv {\hat U(q)}^\dagger h^\mu_{1+i2} {\hat U(q)}={\rm exp}(i{\bf q}\cdot{\bf r})h^\mu_{1+i2},
\eeq
Thus we can see that DCDW supply an extra momentum to modify the momentum conservation in the $\beta$-decay process. It is easily shown that the neutrino emission due to the quark $\beta$-decay is suppressed by the energy-momentum conservation:  the momenta of all the particles should be collinear for reaction (\ref{betad}) for free massless quarks at low temperature. So DCDW catalyzes the reaction (\ref{betad}), which may give rise to a fast cooling of compact stars \cite{muto}.

The symmetry breaking pattern in the DCDW phase is as follows: the original translational symmetry along $z$-axis (the generator is $\hat p_z$) and $U(1)$ subgroup of chiral symmetry (the generator is $Q_5^3$) is broken at the same time, but it is still invariant under a combination of these operations,
\beq
T_{\hat p_z}\times U_{Q_5^3}(1)\rightarrow U_{{\hat p_z}+Q_5^3}(1).
\eeq 

Actually we can see that the single-particle wave function is the eigenfunction of the operator, 
\beq
\exp(i{\hat p}_z a)\exp(-i\gamma_5\tau_3/2\theta)\psi_k(\vr)=\exp(ik_z a)\psi_k(\vr)
\eeq
with $\theta=qz$.
The collective modes ("phasons") have then hybrid properties of phonons and "pions". We can derive the dispersion relation of the collective modes by way of the Ginzburg-Landau theory. If we consider the phase fluctuations, $u(\br,t)$, the free energy should be written as 
\beq
F=F_0(\Delta, q)+\delta F(u(\br,t)),
\eeq
with 
\beq
\delta F=\frac{1}{2}\int 
\left[A(\nabla_z u)^2+B(\nabla_\perp^2 u)^2+C\left(\frac{\partial u}{\partial t}\right)^2\right]d^3 r,
\eeq
like liquid crystal (smectics) \cite{deg}, where the coefficients $A,B,C$ are the function of $\Delta, q$\cite{gru}. 
The Euler-Poisson equation for $u$ then reads,
\beq
C\frac{\partial^2 u}{\partial t^2}-A\frac{\partial^2 u}{\partial z^2}+B(\nabla_\perp^2)^2 u=0.
\eeq 
The plane-wave solution, $u=u_0 e^{i(\omega t-\vk\cdot\br)}$, is obtained 
with the anisotropic dispersion relation,
\beq
\omega^2=(Ak_z^2+Bk_\perp^4)/C.
\eeq
Note that the dispersion relation exhibits a hybrid nature of type I and II Nambu-Goldstone bosons \cite{nie}: 
 $\omega\propto k_z$ for the longitudinal excitation with $k_\perp=0$, 
while $\omega\propto k_\perp^2$ for the transverse excitation with $k_z=0$.

\subsection{Magnetic aspect of DCDW}

Here we reveal the magnetic aspect of DCDW \cite{DCDW}.
Since explicit form of the spinor can be analytically obtained within the mean-field approximation, 
one can evaluate various expectation values in the DCDW phase. 

By the use of the spinor $\psi_w$ one can evaluate the 
expectation values of bilinear form of gamma matrices $\mathcal{O}$ in a simpler way, 
\beq
\langle {\psi}^\dagger(\br) \mathcal{O} \psi(\br) \rangle 
= \int\frac{d^3p}{(2\pi)^3} \langle \psi_w^\dagger(\vp) e^{i\tau_3 \gamma_5 \vq\cdot \vr/2} \mathcal{O} 
e^{-i\tau_3 \gamma_5 \vq\cdot \vr/2}\psi_w(\vp) \rangle. 
\eeq
Spin expectation value, for instance, $\mathcal{O}=\gamma_0 \gamma_5 \gamma_3 /2\equiv \Sigma_z$, 
vanishes because it is proportional to the stationary condition with respect to wave number $q$, 
\beq
\int\frac{d^3p}{(2\pi)^3} \langle \psi_w^\dagger(\vp) \Sigma_z \psi_w(\vp)\rangle 
\propto \left. \frac{\partial \Omega}{\partial q}\right|_{\rm DCDW}=0. 
\eeq

On the other hand, there is formed a spatial modulation of magnetic moment 
along with DCDW. 
The Gordon decomposition of the gauge coupling term provides 
the magnetic interaction in presence of external gauge field: 
$\frac{Q}{2M^*} \bar{\psi}\sigma_{\mu\nu}\psi F^{\mu\nu}$. 
Only z component of the anomalous magnetic moment remains finite, 
\beqa
&&\langle \bar{\psi}(\br) \sigma_{12} \psi(\br) \rangle =
\int\frac{d^3p}{(2\pi)^3} \langle \psi_w^\dagger(\vp) \gamma_0 \sigma_{12} 
\psi_w(\vp)\rangle \cos({\vq\cdot \vr}),  \nn
&&
\langle \bar{\psi}(\br) \sigma_{23} \psi(\br) \rangle =
\langle \bar{\psi}(\br) \sigma_{31} \psi(\br) \rangle =0, 
\eeqa
where 
\beq
\langle \psi_w^\dagger(\vp) \gamma_0 \sigma_{12} 
\psi_w(\vp)\rangle =\frac{2M^*}{\sqrt{M^{*2}+p_z^2}} \ldk n_+(\vp)-n_-(\vp)\rdk 
\eeq
with $n_\pm$ being the Fermi-Dirac distribution function with energy spectra $\epsilon_\pm$. 
Therefore, the magnetic expectation value reflects the anisotropy 
in momentum distribution of the DCDW phase, and depends on finiteness of the effective mass $M^*$. 

To see a possible consequence of the spatially modulated magnetic order, 
we evaluate its local magnetic-flux density with effective charge $Q=\lk \frac{2}{3}-\frac{1}{3} \rk e$, 
\beq
\Phi=\frac{\it Q}{2M^*} \frac{\langle \bar{\psi}(\br) \sigma_{12} \psi(\br) \rangle}
{\langle \psi^\dagger(\br) \psi(\br) \rangle} 3\rho_B, 
\eeq
where we have evaluated magnetic expectation value par quark, and multiplied by quark density. 
For densities normalized by the normal nuclear density $\rho_B/\rho \sim 3-4$, 
the local flux $\Phi$ is estimated to be $O(10^{16})$ G, 
which well exceeds the critical magnetic field $B_{\rm QED}$. 

It might be interesting to recall that the stable Hartree-Fock solution for the case of the short-range potential is always either the normal paramagnetic state or the uniform ferromagnetic state \cite{her66}. We have considered the zero-range interaction between quarks within the NJL model, but spin density wave emerges in our case.

\subsection{Deformed DCDW} 

We must take into account the symmetry breaking effect or the finite current mass for the extension of the above framework \cite{boe}. It is important for a realistic discussion of the phase transition at moderate densities. Moreover, it may become important to elucidate the appearance of various inhomogeneous phases. The real kink crystal (RKC) can be easily generalized to include massive quarks, and has been studied on the temperature-density plane \cite{nic}. Their results show an interesting change of the domain of the RKC as the current mass is changed; it looks to shrink for larger mass. 

Recently we have shown that the symmetry breaking effect can be taken into account by a variational approach, without spoiling the original features \cite{nov}. 
Using the ansatz for the condensates,
\beqa
\langle{\bar \psi}\psi\rangle&=&\Delta\cos\theta(\vr)\nonumber\\
\langle{\bar \psi}i\gamma_5\tau_3\psi\rangle&=&\Delta\sin\theta(\vr),
\eeqa
represented by the spatially dependent chiral angle $\theta(\vr)$, one must solve the Hartree equation self-consistently in the presence of current-mass term, but it should be a hard task. Instead, we apply approximate methods, like a perturbative method \cite{mae2}. Here we take a variational method, which may give better results than the perturbative one. 
Anyway the symmetry breaking effects should be small in the light of the success of the chiral symmetry approach to various phenomena. So we take into account only the deformation of the chiral angle, leaving the quark wave function unchanged. Then we can see that $\theta$ must satisfy the sine-Gordon equation,
\beq
C_A\frac{d^2\theta(z)}{dz^2}+m_c\Delta\sin\theta(z)=0
\label{dsg}
\eeq
at the leading order of $m_c$. It should be interesting to observe the appearance of the sine-Gordon equation. The relevant solution is then written as 
\beq
\theta=\pi+2{\rm am}\left(m_\pi^* z/k,k\right),
\eeq
in terms of the Jacobian elliptic function, where we introduced the effective pion mass by the relation, $C_Am_\pi^{*2}=-m_c\Delta$, 
and ${\rm am}(\xi, k)$ is the amplitude with modulus $k$.
In the limit, $k\rightarrow 0$ with keeping $m_\pi^*/k=$const.($\equiv q/2$), $\theta\rightarrow \pi+qz$ 
so that we have the DCDW solution again in the chiral limit. On the other hand,
$\theta\rightarrow 4{\rm tan}^{-1}\left[e^{m_\pi^*z}\right]$, in the opposite limit, $k\rightarrow 1$, which is the well-known kink solution. 
Thus our variational ansatz may be regarded as an embedding the sine-Gordon kink crystal in $1+1$ dimension into $1+3$ dimensional quark matter; actually we can easily see that $C_A\rightarrow f_\pi^2$ ($f_\pi$: pion decay constant) and Eq.~(\ref{dsg}) is reduced to the standard sine-Gordon equation in $1+1$ dimension,
\beq
\frac{d^2\theta(z)}{dz^2}-m_\pi^2\sin\theta(z)=0,
\eeq
by way of the Gell-Mann-Oakes-Renner relation, $f_\pi^2m_\pi^2=-m_c\langle {\rm vac}|{\bar q}q|{\rm vac}\rangle$ \cite{schon}. 

\section{Summary and Concluding remarks}

It should be important to study EOS at high-density region from various viewpoints. Here we tried to extract information about the properties of hadron or quark matter and EOS, confronting the magnetic phenomena of pulsars.
We have seen two kinds of the magnetic properties in quark matter: one is spontaneous spin polarization and the other is the spin density wave. The former is similar to the itinerant electrons in QED, while the latter is related to the chiral symmetry of QCD. These discussions may be in a rather primitive level and must be verified by observations or more elaborate theoretical studies toward more realistic description. Comprehensive study of these magnetic properties is also needed to study their competition. 

Energetics or the mechanism of the giant flares observed in some magnetars may give a direct hint about the origin of the magnetic field. Thermal evolution of compact stars should be important for us to get the information about the EOS and properties of high-density matter. Actually the magnetic properties is not directly related to EOS but may manifest through the thermal activities of compact stars.

There have been done many works about color superconductivity (CSC) in high-density matter \cite{csc}. It also has implications on compact star phenomena. So it should be interesting to study the interplay of CSC and magnetic properties in quark matter. In ref.\cite{nak03} we have discussed a coexistence of CSC and ferromagnetic order.

More elaborate studies are necessary for inhomogeneous phases: relations among the various phase should be figured out as well as their properties. For example, there seem many resemblances between pion condensation in hadron matter \cite{pio} and DCDW state. One may find a kind of hadron-quark continuity across the deconfinement transition. 
They may also appear during relativistic heavy-ion collisions. So it is interesting to consider how we can observe them in this context.

\printindex
\end{document}